\documentclass[fleqn,10pt]{wlscirep}
\usepackage[utf8]{inputenc}
\usepackage[T1]{fontenc}
\usepackage{upgreek}
\usepackage{IEEEtrantools}

\title{Raman-free fibered photon-pair source}

\author[1]{Martin Cordier}
\author[2]{Philippe Delaye}
\author[3]{Fr\'ed\'eric G\'erôme}
\author[3]{Fetah Benabid}
\author[1,*]{Isabelle Zaquine}
\affil[1]{LTCI, T\'el\'ecom Paris, Institut Polytechnique de Paris, 91120 Palaiseau, France}
\affil[2]{Laboratoire Charles Fabry, Institut d'Optique Graduate School, CNRS, Universit\'e Paris-Saclay, 91127 Palaiseau cedex, France}
\affil[3]{GPPMM Group, XLIM Research Institute, CNRS UMR 7252, Universit\'e de Limoges, Limoges, France}

\affil[*]{isabelle.zaquine@telecom-paris.fr}
\graphicspath{{images/}}

\begin{abstract}
Raman-scattering noise in silica has been the key obstacle toward the realisation of high quality fiber-based photon-pair sources. Here, we experimentally demonstrate how to get past this limitation by dispersion tailoring a xenon-filled hollow-core photonic crystal fiber. The source operates at room temperature, and is designed to generate Raman-free photon-pairs at useful wavelength ranges, with idler at the telecom, and signal at a visible range. We achieve a coincidence-to-accidentals ratio as high as 2740 combined with an ultra low heralded $g_H^{(2)}(0)$ = 0.002, indicating a very high signal to noise ratio and a negligible multi-photon emission probability. Moreover, by gas-pressure tuning, we demonstrate the control of photon frequencies over a range as large as 13 THz, covering S-C and L telecom band for the idler photon. This work demonstrates that hollow-core photonic crystal fiber is an excellent platform to design high quality photon-pair sources, and could play a driving role in the emerging quantum technology.
\end{abstract}
\begin{document}

\flushbottom
\maketitle
%
%
\thispagestyle{empty}


\noindent Among the numerous platforms that have been tested in quantum information, $\chi^{(2)}$ nonlinear media and, in particular, bulk crystals have been historically the workhorse for photon-pair generation due to the relative ease in terms of experimental setup, their cost and their availability \cite{kwiat1995new}. Four-wave mixing mechanism in $\chi^{(3)}$ media, and in particular waveguides and optical fibers have gained much attention in the recent years because of the new prospects they offer in terms of scalability and integrability. For instance, more than 550 silicon-based photonic components have been recently integrated on a single silicon chip, including 16 identical four-wave mixing photon-pair sources \cite{wang2018multidimensional}. Alternatively, fiber-based photon-pair sources have many advantages; they are easily manufacturable, cost effective, robust, alignment-free and compatible with fiber optical network since the photon-pairs are directly emitted in the fundamental transverse guided mode of a fiber. Furthermore, and unlike silicon-based materials, optical fibers don't suffer from two-photon absorption or free carriers effect \cite{husko2013multi}, and a dispersion that can be readily engineered.  

\begin{figure*}[hbt!]
\centering
\includegraphics[width=\linewidth]{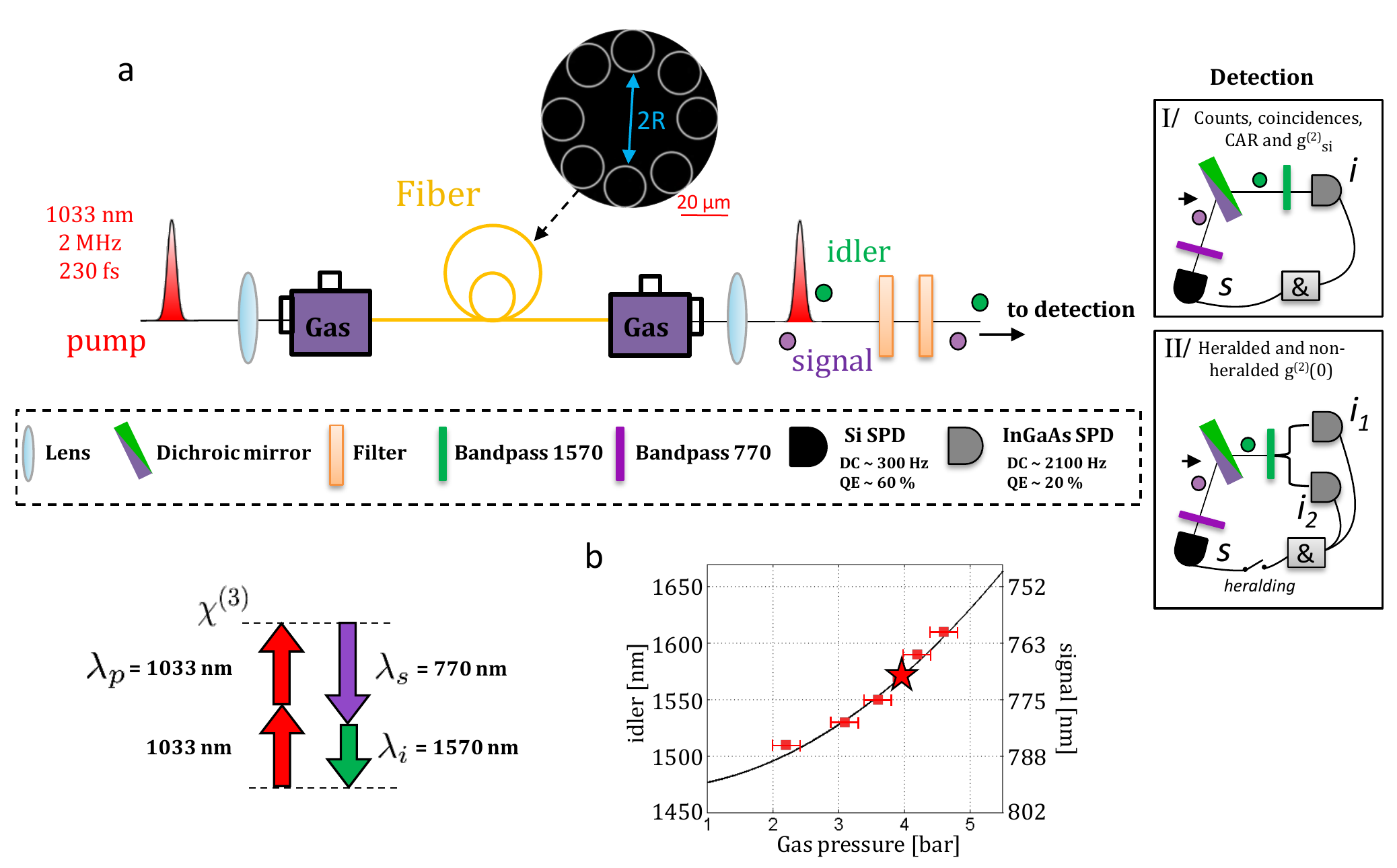}
\caption{\textbf{(a)} Experimental setup. Light pulses from a Yb laser with a duration of 230 fs, repetition rate of 2 MHz and a central wavelength of $\lambda_p = 1033$ nm are coupled to the gas-filled HCPCF. The propagation loss at pump, signal and idler wavelengths are below 0.1 dB/m (see Supplementary note 1.e). Prior to the fiber, a half-wave plate controls the laser polarization. At the fiber output, the signal and idler are separated thanks to a set of 3 dichroic mirrors, together with one broadband filter (shortpass and longpass filter for the signal and idler arm, respectively) and one bandpass filter (10 nm width for the signal and 12 nm for the idler) on each arm. Since signal and idler wavelengths are not in the same frequency range, the detectors are either Si or InGaAs single-photon detectors (both with single mode fiber coupling). (\textbf{b}) Measured central idler frequency as a function of xenon gas pressure. The right axis gives the associated signal frequencies fulfilling energy conservation. The dark line corresponds to the simulated phase-matched frequencies. The red star indicates the experimental configuration. }
\label{fig:setup}
\end{figure*} 

Within the fiber-related endeavors, photon-pairs have been produced in many different architectures encompassing single-mode fiber \cite{hall2009drop, hall2011ultrafast}, dispersion-shifted fiber \cite{liang2006ultra, dyer2009high, medic2010fiber, zhou2013polarization}, birefringent single-mode fiber \cite{smith2009photon, fang2014polarization}, micro/nano-fiber \cite{su2018micro, cui2013generation} and photonic crystal fiber (PCF) \cite{sharping2004quantum, rarity2005photonic, fulconis2007nonclassical, fan2007bright, goldschmidt2008spectrally, halder2009nonclassical, cohen_tailored_2009}. However, the performances of these photon-pair sources are most often plagued by a concomitant nonlinear effect, Raman-scattering (RS).   
Indeed, in addition to the four-wave mixing process leading to photon-pair generation, an interaction between the pump and phonons in the medium results in the generation of scattered photons. Due to the amorphous nature of silica (i.e. the fiber core material), the Raman gain is broadband and continuous \cite{agrawal2007nonlinear}, thus generating a large amount of scattered photons around the pump frequency. In fact, regardless of the photon-pair frequencies, some Raman-scattered photons will be generated at the same frequencies and same polarisation and thus cannot be filtered. For this reason, RS degrades the achievable signal to noise ratio \cite{fiorentino2002all,inoue2004generation}. RS can be reduced by cooling down the fiber to cryogenic temperature at T = 77 K \cite{park2018observation, zhou2013polarization, yang2011characterization, medic2010fiber, hall2009drop}, T = 4 K \cite{dyer2009high} or even 2 K \cite{zhu2016fiber}, which does not suppress it completely, and adds a layer of experimental difficulty making it much less practical for applications and limiting integration possibilities. An alternative solution, compatible with room temperature operation, is to generate the photon-pair with a large spectral separation from the pump. The Raman gain spectrum extends as far as 30 THz from the pump frequency with a peak at 13.1 THz \cite{stolen1972raman}. Past this region, the Raman noise is less detrimental. This approach has been mostly investigated in PCF \cite{francis2017fibre, francis2016all, ling2009mode, fan2007broadband, fulconis2007nonclassical, rarity2005photonic} since the versatility of its design allows the generation of photon-pairs with a large detuning from the pump wavelength. However, higher order RS processes still corrupt the parametric photon-pair generation, as described in \cite{rarity2005photonic, harvey2003scalar,cui2013generation}. This results from an interaction with multiple phonons, spreading the RS noise to an even larger separation from the pump. 
An ideal solution consists in changing the propagation medium (silica) to a nonlinear Raman-inactive medium. 
Hollow-core photonic crystal fibers (HCPCF) allow to guide light into a gas or a liquid with a negligible optical overlap with the glass. With Inhibited-Coupling guiding HCPCF (IC-HCPCF) \cite{couny2007generation}, this overlap can be as low as $10^{-6}$. Several  demonstrations have shown the ability to explore Raman-free nonlinear optics \cite{finger2015raman, azhar2013raman, lynch2013supercritical}. However, despite the ubiquitous usage of HCPCF in fields as large and diverse as nonlinear optic, metrology, atom optics, plasma physics, etc  \cite{russell2014hollow, debord2019hollow}, their potential for quantum information has only recently become apparent. Liquid filled HCPCF have been used to explore Raman-tailored photon-pair generation \cite{Cordier:17, barbier2015spontaneous} which consists of optimising the phase-matching frequency ranges in order to avoid an overlap with the discrete Raman-lines of the liquids. Finger \textit{et al}. \cite{finger2015raman, finger2017characterization} use an IC-HCPCF filled with a noble gas (argon) in order to generate Raman-free squeezed-vacuum bright twin-beam.
In this study, we use a gas-filled IC-HCPCF architecture to demonstrate, for the first time, the possibility of Raman-free spontaneous photon-pair generation through four-wave mixing inside a fiber. The combination of this result with the versatility of this IC-HCPCF platform in tuning and engineering the spectral correlation between the two photons of the pairs opens a new technological path to low cost and robust quantum communication networks. 


\section*{Results}
\noindent\textbf{Fiber design.} The source is a hollow-core fiber filled with xenon gas. Figure \ref{fig:setup} shows the fabricated fiber consisting of 8 silica tubes encircling the core area (radius R =  $22$ $\upmu$m). The strategy to reduce the RS noise to a negligible level is threefold.
Firstly, since light propagates into a hollow-core, only a minute fraction of the optical spatial mode overlaps with silica ($< 10^{-5}$ according to our simulations). Secondly, filling the core with a noble gas ensures that the optical guided mode interacts with a Raman-scattering inactive medium. And thirdly, to mitigate the effect of the tiny overlap with silica, the fiber design was chosen to allow far-detuned phase-matching where signal and idler photon frequencies are separated by 99.3 THz from the pump \cite{cordier2019active}. 
With such large detuning, the photons are created well above the most significant Raman line located at 13.1 THz. Only higher order RS interactions involving more than 5 phonons lead to such detuning \cite{cui2013generation,rarity2005photonic}. Such a far-detuned phase-matching has also the beneficial effect of a negligible pump diffusion and allows the use of lossless dichroic mirrors and broadband bandpass filters to separate the emitted photons from the pump. By eliminating the RS, the main objective of this strategy is to generate fibered photon-pairs with the highest possible signal to noise ratio. Finally, the fiber dispersion has been designed such that the phase-matching configuration gives a signal at 760-800 nm wavelength range, which is compatible with Rubidium or Cesium absorption spectral locations, and lies in the range of Silicon single photon detector, and an idler at telecom wavelength (See Supplementary note 1.d).
Moreover, as previously demonstrated \cite{cordier2019active}, such IC-HCPCF allows to engineer the spectral correlations. Here we have chosen a configuration that minimizes spectral correlations \cite{garay-palmett_photon_2007}. This allows, when using the signal photon to herald the idler, to obtain a telecom wavelength single photon with high state purity.

Figure \ref{fig:setup}.b illustrates the source tunability by showing the measured signal and idler wavelengths when the gas pressure is varied from 2 to 5 bar \cite{cordier2019active}. The results show that within this short pressure range, the idler covers both S-C and L telecom band, when the fiber is pumped with laser set at 1033 nm. The star symbol indicates our operating experimental parameters (i.e. a gas-pressure of 4 bar, a signal wavelength at $\lambda_s = 770$ nm and the idler at $\lambda_i = 1570$ nm) set for the heralded single photon source detailed below. \\

\noindent\textbf{Photon-pair source.}
The set-up to characterize the performances of the source is described in Fig. \ref{fig:setup}.a. Femtosecond laser pulses are injected into the fundamental spatial mode of the gas-filled HCPCF and create co-propagating photon pairs within the fiber core. At the fiber output, the signal and idler photons are filtered and detected individually using standard coincidence correlation scheme. Our analysis relies on the pump power dependence of i) the recorded counts on signal and idler path detectors and ii) peaks of the histograms of time delays between signal and idler detections. 
The recorded counts include parametric photons evolving quadratically with pump power, Raman-scattered photons evolving linearly and the independently measured constant dark counts with respective coefficients $c_x$, $b_x$ and $a_x$ with $x \in$ $\{$s, i$\}$ for signal and idler.
Thus, by fitting the evolution of the count number as a function of the injected pump power $P$, one can evaluate the different contributions:
\begin{IEEEeqnarray}{rCl}
    \centering  N_x = a_x+ b_x P + c_x P^2
\end{IEEEeqnarray}

\begin{figure}[t!]
\centering
\includegraphics[width=\linewidth]{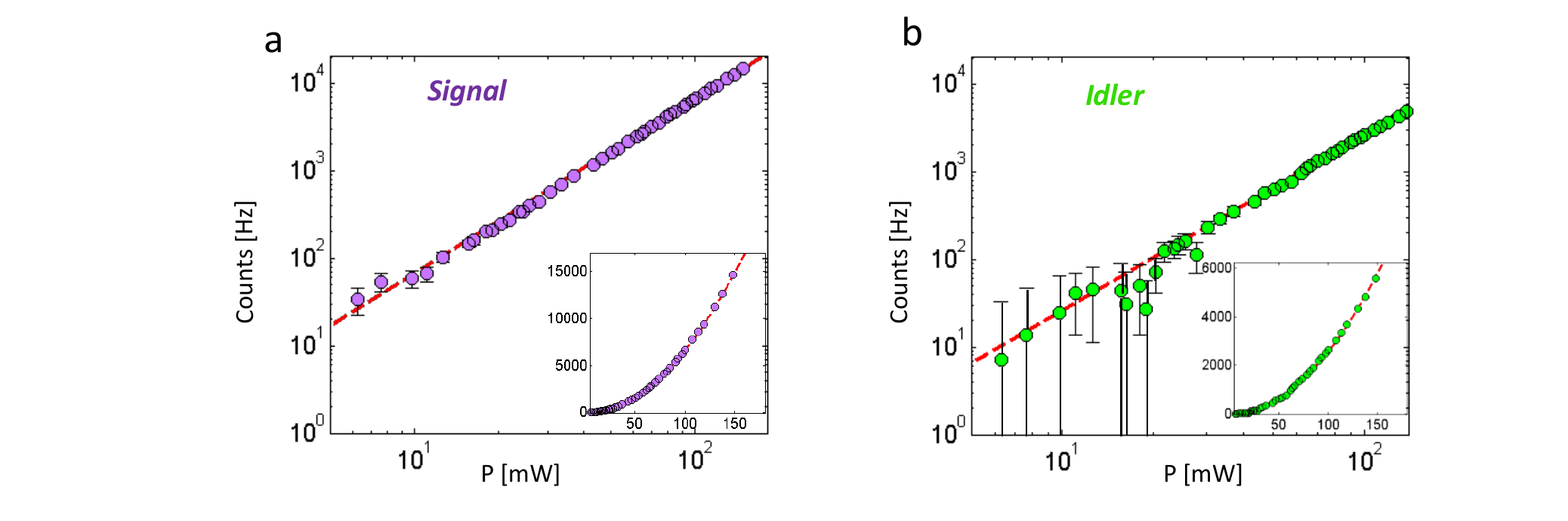}
\caption{(\textbf{a}) Signal and (\textbf{b}) idler counts as a function of the average pump power. The red lines correspond to a pure quadratic fit ($\text{c}_{s/i}$ $P^2$). The dark counts have been subtracted ($a_s = 313$ Hz, $a_i = 2120$ Hz). Fit parameters: $c_s$ = 0.66 $\text{s}^{-1} \text{mW}^{-2}$ and $c_i$ = 0.27 $\text{s}^{-1} \text{mW}^{-2}$. The different single-photon count rates in the two channels are mainly due to the different quantum efficiencies of signal and idler detectors.}
\label{fig:count}
\end{figure}

\begin{figure*}[t!]
\centering
\includegraphics[width=\linewidth]{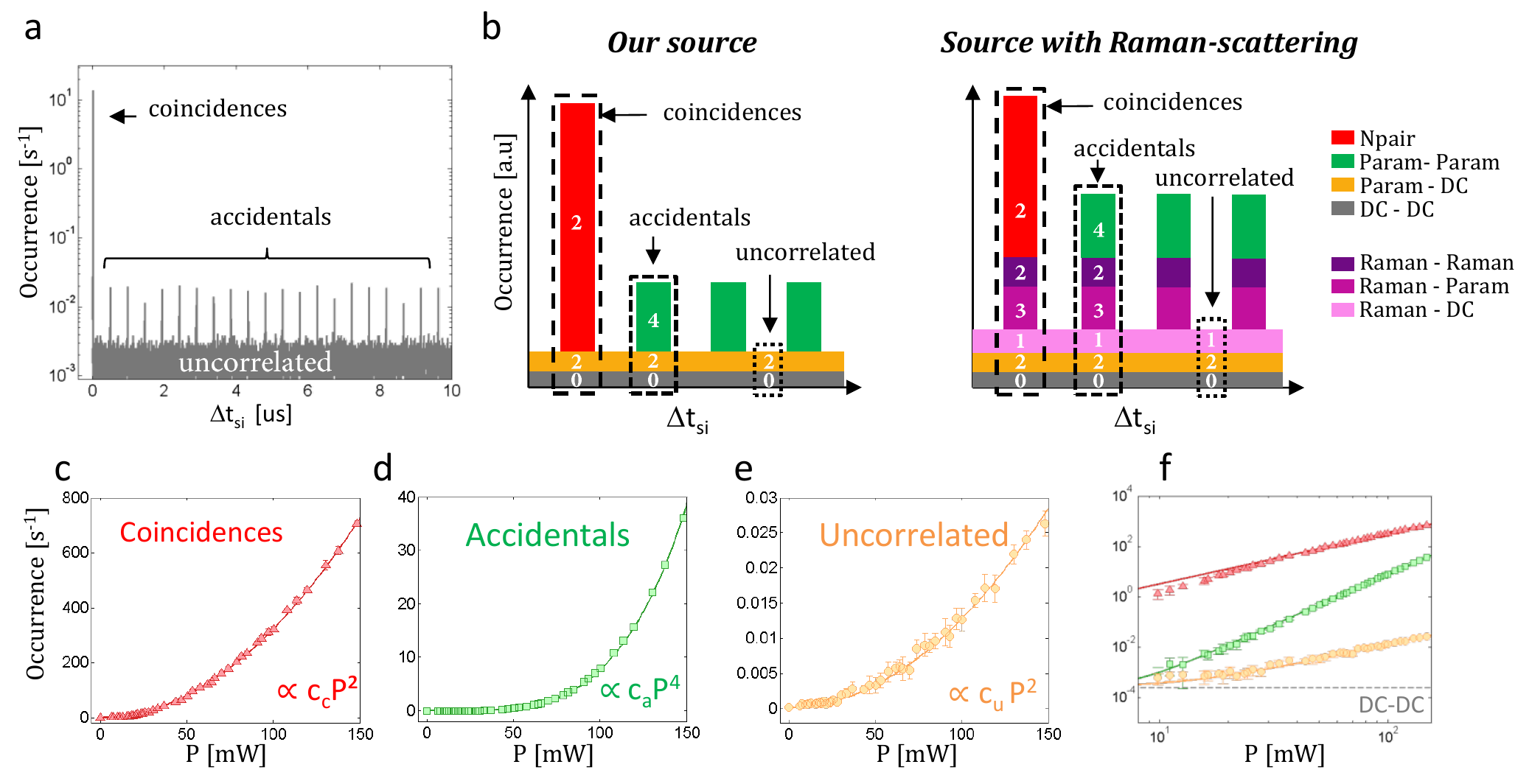}
\caption{ (\textbf{a}) Recorded time delay histogram: occurrences in log scale as a function of the delay $\Delta t_{si}$ between signal and idler detections (P = 30 mW, $t_\text{bin}$ = 1.4 ns). (\textbf{b}) Schematic time delay histograms in arbitrary units with the power dependence (i $\rightarrow P^i$) of the various contributions to the peaks in our source compared to a source with RS. (\textbf{c,d,e}), measured coincidences ($N_\text{coinc}$), accidentals ($N_\text{acc}$) and  uncorrelated ($N_\text{unco}$), as a function of the average pump power in linear and (\textbf{f}) log scale. 
The coloured lines correspond to the polynomial fit with fit parameters: $C_{c}$ = $3.1$ x $10^{-2}$ \, $\text{s}^{-1}\text{mW}^{-2}$, $C_{a}$ =  $7.8$ x $10^{-8}$ \,$\text{s}^{-1}\text{mW}^{-4}$ $C_{u}$ = $1.2$ x $10^{-6}$ \, $\text{s}^{-1}\text{mW}^{-2}$.  }
\label{fig:coinc_acc}
\end{figure*}

\begin{table*}[t!]
\centering
\footnotesize
\caption{ \textbf{Different time histogram events and their characteristics }}
{\renewcommand{\arraystretch}{1}
\renewcommand{\tabcolsep}{0.4cm}
\begin{tabular}{|llll|}
\hline
\textbf{Type of detection events} & \textbf{Denomination} & \textbf{Power dependence}  & \textbf{Position in the histogram} \\
\hline
2 parametric from the same pair  & $N_\text{pair} $ & $P^2$ & Coincidence \\
2 parametric from different pairs & $N_\text{param-param} $ & $P^4$ & Accidentals \\
Parametric and DC & $N_\text{param-DC} $ & $P^2$ & Uncorrelated  \\
DC and DC & $N_\text{DC-DC} $ & Constant & Uncorrelated \\
\hline
\hline
\multicolumn{4}{|c|}{Additional events in presence of Raman-scattering}\\
\hline
Raman and parametric & $N_\text{param-Raman}$ & $P^3$ & Coincidence and accidentals  \\
Raman and Raman& $N_\text{Raman-Raman}$ & $P^2$ & Coincidence and accidentals  \\
Raman and DC & $N_\text{Raman-DC} $ & $P$ & Uncorrelated \\
\hline
 \end{tabular}}
  \label{Table:coicn}
    \end{table*}

The count rate of signal and idler photons as a function of the average pump power is shown in Fig. \ref{fig:count}. 
The perfect match (see Supplementary Note 2.a) with a quadratic polynomial ($\propto P^2$), over more than one decade, is a first element of proof of photon-pair generation and of the negligible RS. Furthermore, the number of detected signal and idler counts can then be simply written: $N_{s/i} = \text{DC}_{s/i} + T_{s/i} \,\eta \, P^2$ where $ \text{DC}_{s/i}$ designates the dark count, $\eta$ the generation efficiency, $P$ the pump power and $T_{s/i}$ the overall transmission in the signal and idler paths including detector quantum efficiencies (typically $T_s \approx 18 \%$ and $T_i \approx 6 \%$ essentially limited by detector quantum efficiencies and coupling efficiency in the single mode fibers connected to the detectors). 



Figure \ref{fig:coinc_acc}.a shows an example of a recorded histogram depicting the number of occurrences within a 10 $\upmu s$ window, as a function of the delay $\Delta t_{si}$ between signal and idler detections.
Three types of events can be distinguished through their time delay:

\begin{itemize}
\item[-] \textit{Coincidences} ($ N_\text{coinc}$) i.e. simultaneous detection from the signal and idler paths mostly quantifies signal and idler photons from the same pair $N_\text{pair}$. A small fraction of this main peak is due to "false coincidences", which can result from noise photons (Raman or DC).

\item[-] \textit{Accidentals} ($N_\text{acc}$) from the secondary peaks correspond to either parametric  ($N_\text{param-param}$) or Raman-scattered ($N_{param-Raman}$ and $N_{Raman-Raman}$) photons generated by different pump pulses. These peaks are separated by multiples of the $500$ ns period corresponding to the $2$ MHz repetition rate of the laser. 

\item[-] \textit{Uncorrelated} events ($N_\text{unco}$) are not temporally correlated to the pump pulses. They involve a dark count on the first detector and parametric ($N_{param-DC}$) or Raman photon ($N_{Raman-DC}$), or another dark count ($N_{DC-DC}$) on the second detector and they are distributed on all the bins of the histogram. 
\end{itemize}
The various contributions to the different types of events can be evaluated through their different pump power dependence and their position in the temporal histogram are described in Table \ref{Table:coicn} and Fig. \ref{fig:coinc_acc}.b. Consequently, in our source, the clear quadratic power dependence of the coincidences (Fig. \ref{fig:coinc_acc}.c) and uncorrelated events (Fig. \ref{fig:coinc_acc}.e) as well as the net power fourth dependence of the accidentals (Fig. \ref{fig:coinc_acc}.d) are a second strong signature of negligible RS noise (see also Fig. \ref{fig:coinc_acc}.f for a comparison in log scale). Indeed, in the case of RS, which is not detected here, odd power polynomial terms in pump power would have appeared (see Fig. \ref{fig:coinc_acc}.b and Supplementary Note 2.a for detailed fit parameters). The further characterizations presented hereafter confirm this analysis:


A standard measure of the signal to noise ratio of a photon-pair source is the coincidence-to-accidental ratio (CAR):
\begin{IEEEeqnarray}{rCl}
\text{CAR} = \frac{N_\text{coinc}}{N_\text{acc}} 
\end{IEEEeqnarray}

\noindent 
In the regime of negligible dark-counts, the CAR of our source is expected to decrease as $1/P^2$ as the numerator increases quadratically with pump power ($N_\text{coinc} \approx N_\text{pair}$) whereas the denominator has a power fourth dependence with power ($N_\text{acc} \approx N_\text{param-param}$). It is important not to associate $N_\text{param-param}$ with the detection of multiple-pairs within one pump pulse. Such events are actually negligible in first approximation as the number of generated photon per pulse is of the order of $10^{-3}$. $N_\text{param-param}$ rather corresponds to the detection of parametric photons from different pump pulses. On the time-delay histogram, this term therefore does not appear in the coincidence peak but only in the accidentals. This contribution is related to the effect of losses in the setup and not to an intrinsic noise in the source (see Supplementary note 2.c). Figure \ref{fig:CAR} shows the measured CAR of the source as a function of the average pump power. 
This decreasing CAR with pump power results in the usual trade off between brightness and signal to noise ratio. The CAR reaches a maximum of 2740 which is, to our knowledge, the best CAR reported in any fiber architecture. This high CAR, obtained at room temperature is a direct consequence of the drastic reduction of the RS noise. This value is currently only limited by the high level of dark-counts ($\text{DC}_i \approx 2$ kHz / $\text{DC}_s \approx 0.3$ kHz) and the low quantum efficiency ($\text{QE}_i \approx 0.2$ / $\text{QE}_s \approx 0.6$) of the used detectors, paving the way to even higher performances (see Supplementary note 2.b). 
\begin{figure*}[t]
\centering
\includegraphics[width=\linewidth]{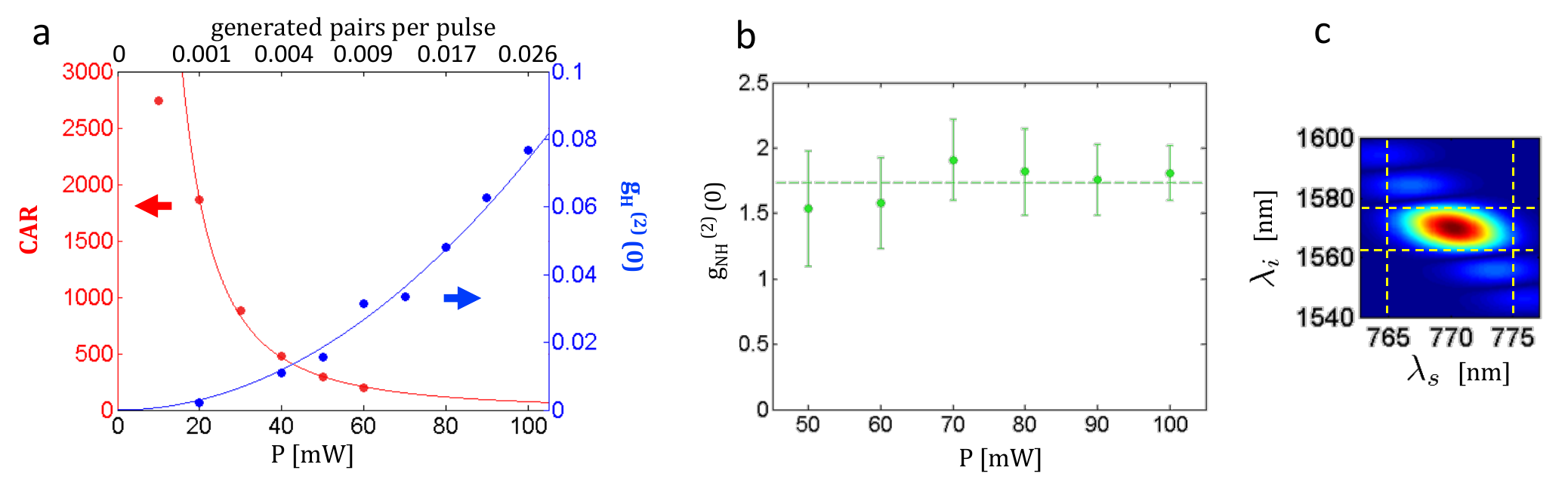}
\caption{(\textbf{a}) CAR and heralded second order cross-correlation as a function of the average pump power. The best CAR equals 2740. The integration time is 1 hour. The measured CAR at lowest power (10 mW) deviates from the trend as the accidentals peaks become saturated by the uncorrelated noise. The red and blue curves are a fit of the form $\propto 1 / P^2$ and $P^2$ respectively.  (\textbf{b}) Non-heralded second order coherence. (\textbf{c}) Simulated joint spectral amplitude. The dashed lines give the width of the bandpass filters used in the experiment. }
\label{fig:CAR}
\end{figure*}

In the absence of RS, the generation efficiency $\eta$ can be easily inferred from the power dependence of the second-order correlation function defined as the ratio of coincidences and the product of signal and idler counts \cite{spring2013chip}: 
\begin{IEEEeqnarray}{rCl}
g^{(2)}_{si}(0) = \frac{N_\text{coinc}}{(N_s-DC_s) (N_i-DC_i) } \approx \frac{1}{\eta \, P^2 } 
\end{IEEEeqnarray}
which does not require the knowledge of the overall loss and detection efficiency. We obtain $\eta \approx 2.6 \, .10^{-6} \, \text{mW}^{-2}$/pulse. Considering the power range used in our experiments, this gives a generation rate that varies from $2.6 \, 10^{-4}$ to $2.6 \, 10^{-2}$ per pulse, meaning a photon-pair rate in between 0.5 to 52 kHz.  
It is noteworthy that the low brightness figure of this first demonstration, could be improved by 4 orders of magnitude with an optimized design with realistic parameters: the nonlinearity could be increased by using a smaller fiber core radius; the gain would then be further enhanced by the higher gas pressure required to obtain the same phase-matching condition as nonlinearity also increases with gas pressure (Supplementary Note 1.c). The fiber length could also be increased as well as the repetition rate of the pump laser (2 MHz) that is much lower than the typical 80-100 MHz of many reported sources. Let us stress that the brightness figures depend very strongly on the measurement conditions: pump power, repetition rate and pulse duration, signal and idler detection frequency bandwidths, detector quantum efficiency and dead time.

As a conclusion of this photon-pair source characterization, we have achieved the first spontaneous photon-pair generation in a gas-filled HCPCF and, as a consequence of the negligible RS noise, the highest reported CAR in a fibered source. \\

\noindent\textbf{Heralded single-photon source.}
This source can be used as a heralded single photon source at telecom wavelength by using the signal photon at 770 nm to herald the idler. The quality of the resulting source can be derived from the heralded second order coherence function $g^{(2)}_H$. It is measured by adding a beamsplitter on the idler arm (see Fig. \ref{fig:setup} inset II) and comparing threefold and twofold coincidences between the signal and the two idler single photon detectors \cite{spring2013chip, mauerer2009colors}: 
\begin{IEEEeqnarray}{rCl}
g^{(2)}_H(0) = \frac{N_{s,i_1,i_2} N_s}{N_{s,i_1} N_{s,i_2}}
\end{IEEEeqnarray}
with $N_{x,y,z}$, $N_{x,y}$, $N_{x}$ the number of threefold coincidences, twofold coincidences and counts, respectively, with $\{x,y,z\}$ = $\{s,i_1,i_2\}$. \\
$g^{(2)}_H(0) = 0$ for a perfect single photon source \cite{caspani2017integrated}. A non-zero value reveals the presence of multi-photon generation or any source of noise (RS, residual pump, etc). Figure \ref{fig:CAR} shows the measured heralded $g^{(2)}_H(0)$ as a function of pump power. At 20 mW, the value reaches $g^{(2)}_H(0) = 0.002 \pm 0.001$, meaning that the probability of getting more than one photon per pulse is 7 orders of magnitude lower than the one of having a single photon \cite{migdall2013single}. To our knowledge, this is the lowest $g^{(2)}_H(0)$ obtained in a fiber architecture. This high performance in terms of source quality is also a consequence of the \textit{Raman-free} feature of our source. 
Moreover, the measured $g^{(2)}_H(0) $ increases quadratically with pump power (blue curve) as expected \cite{yang2011characterization}, whereas it would saturate to a linear dependence in the presence of RS noise \cite{spring2013chip}. 

\begin{figure}[t]
\centering
\includegraphics[width=0.45\linewidth]{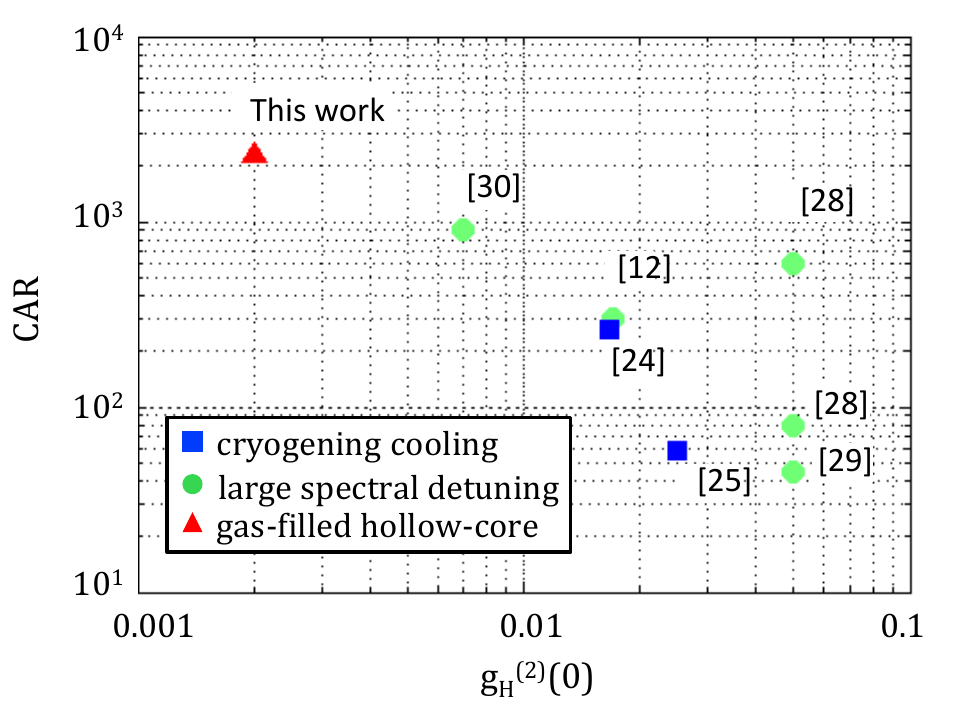}
\caption{Comparison of state of the art fiber photon-pair sources. The corresponding reference is indicated in brackets.  }

\label{fig:etatArt}
\end{figure}
The heralding efficiency, i.e. the probability that an idler photon is present given the detection of its twin, is also a very important characteristic of the heralded photon source. We measure a value of $\eta_\text{herald} \approx 27 \%$ which is comparable with other reported fiber sources \cite{smith2016double, su2018micro, smith2009photon}. 

Another fundamental parameter is the purity of the heralded telecom photon. The fiber design was chosen to exhibit near spectral factorability of the photon-pair state, as demonstrated in \cite{cordier2019active}. The singular-value decomposition of the simulated joint spectral amplitude (Fig. \ref{fig:CAR}.c) provides an upper bound of the purity of the idler photon: $P_i < 90 \%$ \cite{law2000continuous}. This upper bound assumes a flat spectral phase and a perfectly homogeneous fiber. The purity can be derived experimentally from the non-heralded second-order coherence $g_{NH}^{(2)}(0)$ which is measured with our setup by ignoring the heralding signal (see Fig. \ref{fig:setup} inset II) and is given by \cite{spring2013chip}: 
\begin{IEEEeqnarray}{rCl}
g_{NH}^{(2)}(0)= \frac{N_{i_1,i_2} R_p}{N_{i_1} N_{i_2}} = 1 + P_i
\end{IEEEeqnarray}
with $R_p$ the repetition rate of the pump laser and $P_i$ the purity of the idler photon. 

\noindent The $g_{NH}^{(2)}(0)$ of the heralded telecom photon is shown in Fig. \ref{fig:CAR}.b. The value is rather constant over the measured power range with a mean value of 1.74 $\pm$ 0.32 and yields a maximum of 1.91 $\pm$ 0.31.  The high purity $P_i \approx 74 \%$ of the telecom photon obtained without resorting to narrowband spectral filtering is an experimental proof of the spectral correlation engineering possibilities of the gas-filled HCPCF platform.


\section*{Discussion}
\noindent We have demonstrated the first tunable photon-pair source based on a gas-filled hollow-core photonic crystal fiber. Filled with a monoatomic gas, this architecture allows Raman-free nonlinear optics within the fiber at room temperature. As a consequence, the source exhibits outstanding performances in terms of signal to noise ratio, with state of the art CAR and heralded $g_H^{(2)}(0)$, despite the low performances of our detectors. This first design already outperforms other well-established strategies to reduce Raman-scattering noise as shown in Fig. \ref{fig:etatArt}. The control of the two photon frequencies over a large range through gas pressure tuning, thus enabling for choosing specific telecom channel or atomic transitions is also a remarkable feature of our source.\\
Furthermore, we show that the source can also be used as a fibered heralded single photon source at telecom wavelength with good heralding efficiency and purity showing that spectral correlation engineering of the photon-pairs has effectively allowed the generation of factorable pair states. 
This first demonstration shows that this new platform of gas-core photonic crystal fibers is very promising and could provide versatile sources that can be designed to satisfy the requirements of many quantum information applications.


\section*{Acknowledgments}

\noindent We thank M. G. Raymer  for useful discussions, GLO photonics for lending some of the equipment used in this experiment, and the GPPMM team for the fiber fabrication. 
This research was supported by IDEX Paris-Saclay (ANR-11-IDEX-0003-02); Labex SigmaLim; Region Nouvelle Aquitaine.

\bibliography{sample}






\end{document}